\documentclass{ere04}

\usepackage{amsmath,amssymb}

\usepackage{graphicx}    

\def\beq{\begin{equation}}
\def\eeq{\end{equation}}

\def\bea{\begin{eqnarray}}
\def\eea{\end{eqnarray}}
\def\ba{\begin{array}}                  
\def\ea{\end{array}}

\def\e{\epsilon}

\begin{document}

\title*{Second order perturbations of rotating bodies in equilibrium;
the exterior vacuum problem.}

\author{M.A.H. MacCallum\inst{1}\and M. Mars\inst{2} \and R. Vera\inst{3}}
\institute{ School of Mathematical Sciences,
Queen Mary, University of London,
United Kingdom
\texttt{M.A.H.MacCallum@qmul.ac.uk}
\and Facultad de Ciencias, Universidad de Salamanca, Spain.
\texttt{marc@usal.es}
\and School of Mathematical Sciences, Dublin City University, Dublin 9,
Ireland
\texttt{raul.vera@dcu.ie}}

%
%
\maketitle

We study the exterior vacuum problem for first and second order stationary
and axially
symmetric perturbations of static bodies. The boundary conditions and their
compatibility for the existence of an asymptotically flat exterior solution
are discussed.

\section{Introduction}
\label{sec:1}

Finding global models for rotating objects in general relativity has
proven to be extremely difficult, even 
for axially symmetric configurations in equilibrium.
So far there are no known explicit global models except for spherical stars
(hence non-rotating) or the
Neugebauer and Meinel disc of dust \cite{NeugebauerMeinel92}, where the matter
source is encoded as suitable jumps on the first derivatives of the metric,
which is otherwise vacuum everywhere. 

Since a proper understanding of rotating objects in
equilibrium within the context of general relativity is obviously
fundamental for many astrophysical situations, alternative methods
have been developed over the years.  The two most important ones are,
without doubt, the use of numerical methods and perturbation
theory. The former can handle fully relativistic situations with
intense gravitational fields and high velocities and has produced a
large variety of very interesting results. However, the fact that
closed expressions are never found in this setting leaves plenty of
room for other approaches, like for instance perturbation theory. Its
field of applicability is restricted to slowly rotating stars (so that
the perturbation parameter can be taken to be the maximum angular
velocity of the star, for instance) and also a wealth of results have
been obtained (see e.g. \cite{LR} and references therein).

This short note is a summary of a longer paper \cite{MMV} by the same
authors where many more results and their proofs can be found. The
object of this contribution is to explain in a few words the main
motivation of this work, present the main result and give some
indications of how it can be proven by analyzing the simplest possible
situation.

The aim of this work is to study perturbation theory of rotating stars from a different perspective
than normally done. We want to consider slowly rotating bodies with an arbitrary matter content and we 
wish to concentrate on the effect that the rotation has in the vacuum outside region. Thus, our
background spacetime is composed of a static object with some non-vanishing energy-momentum tensor
(typically a perfect fluid, but not necessarily). Although we have in mind the case when the static body
is spherically symmetric, which is the physically most relevant one, all our results hold also
for axially symmetric backgrounds.
We assume, as it is usual
in this context, that the object 
has a sharp boundary  and that there is no matter outside it, so that
the metric exterior to the body is vacuum. The metrics inside and outside the body must satisfy
certain junction conditions (see \cite{MarsSenovilla93} for a detailed account) in order to 
produce a well-defined spacetime. Given this background, we want to perturb it {\it
arbitrarily} in the interior except for the restriction that the object is still in
equilibrium and has axial symmetry (i.e. the perturbed metric is stationary and 
axially symmetric). Furthermore we want to do perturbations up to second order in perturbation
theory. The necessity of going to second order comes  from the fundamental results obtained
in the seminal paper by Hartle \cite{Hartle67} where rigidly rotating perturbations of
spherically symmetric and static perfect fluids  were analyzed and where it was found that
to first order rotation only affects the $, \phi$ crossed term in the metric (more technically, it
only produces so-called axial or odd perturbations) without modifying, for instance, the
spherical shape of the object, while second order perturbations, already affect the
shape of the body and the rest of the metric components.

In Hartle's paper some heuristic arguments were used at some places, specially
regarding the matching procedure of the perturbed metrics. One of the aims
of our paper is to set up a proper theoretical framework so that, in a future work, we can
make rigorous all the arguments used by Hartle. This is important insofar Hartle's
paper has served as the basis of many developments in perturbation theory of rotating objects
over the years.

\section{Brief summary of stationary and axially symmetric rotating bodies}

A spacetime describing a stationary and axially symmetric rotating
body with a boundary surface is composed by two regions; one inside
the body solving the Einstein field equations with matter, and another
outside the body solving the vacuum field equations and being
asymptotically flat (because we are dealing with an isolated
body). Furthermore the two metrics must fulfill the so-called matching
conditions on the boundary of the body, i.e. on a timelike, stationary
and axially symmetric hypersurface $\Sigma$. The vacuum field
equations outside the body can be written in terms of a coupled system
of non-linear elliptic PDE for two scalars $U$ and $\Omega$ called the
``norm'' and the ``twist'' potentials respectively and which are
defined in terms of the unique stationary Killing vector $\vec{\xi}$
which is unit at infinity. The twist potential is defined only up to
an arbitrary additive constant which is fixed by demanding $\Omega
\rightarrow 0$ at infinity. With this choice, $\Omega$ vanishes if and
only if the spacetime is static, hence the twist potential determines
whether the body is rotating or not. In the vacuum region, there exist
local coordinates $\{t, \phi, \rho, z \}$ called Weyl coordinates
which are adapted to the stationary and axial Killing vectors and
which locate the axis of symmetry at $\rho=0$. This coordinate system
is defined uniquely except for an additive constant in $z$, which can
be fixed whenever the spacetime has an equatorial plane by choosing
$z=0$ on the equator. We will assume that the Weyl coordinates exist
also globally in the vacuum exterior region (see \cite{Raul05} for
global existence results of the Weyl coordinate system). In this
setting $U$ and $\Omega$ are functions of $\rho$ and $z$ alone and
satisfy the so-called Ernst equations, which read
\begin{equation} \label{eq:ernst}
  \begin{array}[c]{l}
    \displaystyle{\triangle_{\gamma} U+\frac{1}{2}e^{-4U}
    \left(\mathrm{d} \Omega,\mathrm{d} \Omega\right)_\gamma=0},\\
    \displaystyle{\triangle_{\gamma}\Omega-4\left(\mathrm{d} \Omega,\mathrm{d} 
U\right)_\gamma =0},
  \end{array}
\end{equation}
where $\triangle_{\gamma}$ is the Laplacian of the flat metric $\gamma
= d\rho^2 + dz^2 + \rho^2 d\phi^2$ and $( \, , \,)_{\gamma}$ denotes
scalar product with respect to it.  The asymptotic flatness condition
demands $U = 1 - M /r + O (r^{-2})$, $\Omega = -2 z J /r^3 +
O(r^{-3})$, for some constants $M,J$ and where $r^2 = \rho^2 +
z^2$. The boundary of the rotating body as seen from the exterior
region will be denoted by $\Sigma^E$ and will be defined by two
functions $\rho = \rho (\mu)$, $z = z (\mu)$. If the metric inside the
body is assumed to be known, then the matching conditions can be seen
\cite{MarsSenovilla98} (see \cite{conv} for the complete generalisation)
to be equivalent to the following data: (i) the
explicit form of the matching hypersurface in Weyl coordinates,
i.e. the functions $\rho(\mu), z(\mu)$ and (ii) the values of $U$ and
$\Omega$ (the latter except for an additive constant) {\it together
with their normal derivatives} on $\Sigma^E$. Notice that the Ernst
equations are elliptic, which means that appropriate boundary data
consist of fixing the value of the functions {\it or} their normal
derivatives on the boundary (or perhaps a combination of both), but
never the full values of the functions {\it and} their normal
derivatives. In more technical terms the boundary conditions are of
Cauchy type, which is unsuitable for elliptic problems. This property
reflects the fact that given an arbitrary metric describing a rotating
body in equilibrium, in general there will {\it not} exist a vacuum
solution matching with the given metric and extending to infinity in
an asymptotically flat manner. The problem of finding a global model
of a rotating object is truly global in nature and cannot be broken
into an interior and exterior problem without paying some price. In
our case this translates into the necessity of dealing with an
overdetermined boundary value problem for an elliptic system.  Thus,
one has to worry about existence of the solutions, i.e. which are the
restrictions that the boundary data (and hence the interior metric)
must satisfy so that they truly represent an isolated rotating body.
Let us stress here that uniqueness is a much simpler problem which can
be solved in full generality \cite{MarsSenovilla98}.  With regard to
existence, there are results involving an infinite set of
compatibility conditions on the boundary data which are necessary for
existence to hold \cite{MarsERE}. Whether they are also sufficient is
still an open problem.

\section{First and second order perturbations of the exterior region}

After this brief reminder of the 
non-linear case, let us move into perturbation theory. 
Calling $\e$ the perturbation parameter, we consider
a one-parameter family of spacetimes depending on $\e$. 
Since we take every element in the family to be
a stationary and axially symmetric spacetime, vacuum outside some
spatially compact region (defining the rotating body) and
asymptotically flat, we have two families of functions $U_\e (\rho,z)$,
$\Omega_\e (\rho,z)$, which for all $\e$ satisfy the Ernst equations
(\ref{eq:ernst}). Defining $U' \equiv \partial_{\e} U_\e |_{\e=0}$,
$U'' \equiv \partial_{\e} \partial_{\e} U_\e |_{\e=0}$ and similarly
for $\Omega'$ and $\Omega''$, and using now that the background
spacetime is static, i.e. that $\Omega_\e |_{\e=0} =0$, we find the
first and second order perturbed field equations
\begin{equation}
 \left .  \begin{array}[c]{l}
    \displaystyle{\triangle_{\gamma}U'_0 =0}\\
    \displaystyle{\triangle_{\gamma}\Omega'_0-
      4\left(\mathrm{d} \Omega'_0,\mathrm{d} U_0\right)_\gamma=0}
  \end{array} \right \}, \quad
\left . 
  \begin{array}[c]{l}
    \displaystyle{\triangle_{\gamma}U''_0+e^{-4U_0}
      \left(\mathrm{d} \Omega'_0,\mathrm{d} \Omega'_0\right)_\gamma=0}\\
    \displaystyle{\triangle_{\gamma}\Omega''_0-8
      \left(\mathrm{d} \Omega'_0,\mathrm{d} U'_0\right)_\gamma
      -4\left(\mathrm{d} \Omega''_0,\mathrm{d} U_0\right)_\gamma=0}
  \end{array} \right \}, \label{eq:ernstper}
\end{equation}
simply by taking first and second partial derivatives of
(\ref{eq:ernst}) with respect to $\e$ and evaluating the result at
$\e=0$.

Notice that we are performing the perturbation (i.e. derivative with
respect to $\e$) by considering $\rho,z$ as independent of
$\e$. This entails a suitable identification between the different
spacetimes for different $\e$. Such an identification must always be
made in order to define metric perturbations. However, the
identification is not unique, as we could perform an arbitrary
diffeomorphism on each element of the family $V_\e$ of spacetimes and
the diffeomorphisms can obviously depend on $\e$. This freedom in the
identification implies the gauge freedom which is inherent to
perturbation theory in general relativity (and indeed in any geometric
theory). Thus, when we take perturbations by performing derivatives
with respect to $\e$ with fixed $\{\rho,z\}$ we are effectively
fixing the gauge.

We should consider now which is the domain where the equations
(\ref{eq:ernstper}) hold and what kind of boundary data need to be
fulfilled.  Given the interior family of metrics, the matching
conditions fix (for every $\epsilon$) two functions $\rho_{\e}(\mu)$
and $z_{\e}(\mu)$ defining the matching hypersurface (i.e. the
boundary of the body) for every $\e$.  They can also be seen as
two-surfaces defined in Euclidean 3-space $\mathbb{E}^3$ with the flat
metric $\gamma$ written in cylindrical coordinates.  If $\e$ is close
enough to $\e=0$ we have a family of axially symmetric surfaces
$\Sigma_\e$ in $\mathbb{E}^3$, all of them diffeomorphic to the surface
corresponding to the static background, which will be denoted by
$\Sigma_0$. We will furthermore assume that the background surface is
diffeomorphic to a sphere, which is physical reasonable (and certainly
true whenever the static background is spherically symmetric).

Let us also choose the range of variation of $\mu$ so that there exist
two fixed values $\mu_S$ and $\mu_N$ such that $\rho_{\e} (\mu_S)=
\rho_{\e} (\mu_N)=0$ for all $\e$, i.e.  that the intersection
points of the surfaces with the axis of symmetry occurs at the same
value of $\mu$.  As explained before, the matching conditions together
with the interior metrics provide us with four functions, which we
denote as $G_{\e}$, $L_{\e}$, $Y_{\e}$ and $W_{\e}$, all
of them defined on $\Sigma_\e$ and such that the boundary values for the
exterior problems are
$U_\e |_{\Sigma_\e} = G_{\e}$, $\vec{n}_{\e} (U_\e) |_{\Sigma_\e} = L_{\e}$,
$\Omega |_{\Sigma_\e} = Y_{\e}$, $\vec{n}_{\e} (\Omega_\e) |_{\Sigma_\e} = 
W_{\e}$, where
$\vec{n}_{\e} \equiv -\dot z_\e\partial_\rho+\dot\rho_\e
\partial_z|_{\Sigma_\e}$ is a vector field orthogonal to $\Sigma_\e$ (dot
denotes derivative with respect to $\mu$). Notice that in these
expressions the right-hand sides are functions of $\mu$ and $\e$
alone, while for the left hand sides they are functions of
$\mathbb{E}^3$ evaluated on a (moving) surface $\Sigma_\e$.  We can now
take derivatives of these expressions with respect to $\epsilon$ (at
$\mu$ constant) in order to obtain the boundary values for the
perturbed functions $U'_0$, $\Omega'_0$, $U''_0$ and $\Omega''_0$ and
their normal derivatives on the unperturbed surface $\Sigma_0$.  In
particular, it follows that the domain where the perturbed Ernst
equations hold coincides exactly with the domain corresponding to the
static background. Moreover, the boundary data of $U'$, $U''$,
$\Omega'$ and $\Omega''$ on the unperturbed hypersurface $\Sigma_0$ can
be explicitly calculated in terms of the interior background metrics
and the interior first and second perturbation metrics alone. The
resulting expressions are long and will not be given here (see
$\cite{MMV}$ for a detailed discussion and explicit
expressions). Having discussed briefly the perturbed exterior problems
and their boundary conditions, we can now address the issue of
existence of the exterior solution fulfilling these boundary data.

\section{Compatibility conditions}
\label{Sectcompatibility}

In the previous section we have seen that, as in 
the non-linear case, the perturbed exterior problem involves an elliptic 
system of equations with Cauchy boundary data. Again this is
an overdetermined system and we should not expect asymptotically flat solutions to exist
for any Cauchy data (i.e. for any interior perturbation). Thus we need to
address the question of what is the set of necessary and sufficient conditions
that the boundary data must satisfy so that solutions exist. Asymptotic flatness
demands $\lim_{r \rightarrow  \infty} U'_0 =
\lim_{r\rightarrow  \infty} \Omega'_0 =
\lim_{r\rightarrow  \infty} U''_0 =
\lim_{r \rightarrow  \infty} \Omega''_0 = 0$. The perturbed Ernst equations can be collectively
written as
\begin{equation}
\triangle_{\hat{\gamma}} u = j, \label{master}
\end{equation}
where  $u=u(\rho,z)$ stands for $U_0$, $U_0'$, etc..., and 
$j=j(\rho,z)$ represents the inhomogeneous terms in the second order 
perturbation equations. The metric $\hat{\gamma}$ corresponds to 
either $\gamma$, for the $U$-equations, or $\tilde{\gamma} \equiv e^{-8 U_0} \gamma$,
for the $\Omega$-equations. The domain 
$(D_0,\hat{\gamma})$ is clearly unbounded because $\hat{\gamma}$
is an asymptotically flat metric. 
Thus, the compatibility conditions for the boundary values of
$U'_0$, $U''_0$, $\Omega'_0$, $\Omega''_0$ can be 
studied as particular cases for the
compatibility conditions of the Cauchy problem for the
general inhomogeneous
Poisson equation (\ref{master}) defined on an unbounded asymptotically flat
region $(D_0,\hat{\gamma})$ with boundary 
$\Sigma_0= \{ \rho = \rho_0 (\mu), z = z_0 (\mu), \phi = \varphi \}$, corresponding to the boundary of
the static background metric.
Furthermore, we will assume that $j$ tends to zero at infinity at least like 
$1/r^4$ (which follows in our case from asymptotic flatness).

In order to give a flavor of why Theorem \ref{casebycase} holds, let
us concentrate on the simplest possible case, i.e. when $\hat{\gamma} =
\gamma$ and the source term $j$ vanishes.  A simple consequence of
Gauss' theorem is the so-called Green identity, which reads: for any
compact domain $K \in D_0$ and any function $\psi$ (both suitably
differentiable)
\begin{equation}
\int_{K}\left(\psi \triangle_{\gamma} u- u\triangle_{\gamma} \psi\right)
\eta_{\gamma}=
\int_{\partial K}\left [ \frac{}{} \psi \vec{n}_{\gamma} ( u ) -
u \vec{n}_{\gamma} ( \psi ) \right] \mathrm{d} S_{\gamma}, 
\label{Green}
\end{equation}
where $\vec{n}_{\gamma}$ is a unit (with respect to $\gamma$) normal
vector pointing outside $K$, $\eta_{\gamma}$ is the volume form of
$(D_0,\gamma)$ and $\mathrm{d} S_{\gamma}$ is the induced surface element of
$\partial K$. We intend to apply this identity to a function $\psi$
that (i) solves the Laplace equation $\triangle_{\gamma} \psi = 0$ on
$D_0$, (ii) admits a $C^1$ extension to $\partial D_0 \equiv \Sigma_0$
and (iii) it decays at infinity in such a way that $r \psi $ is a
bounded function on $D_0$. A function $\psi$ satisfying these three
properties is called a {\it regular $\gamma$-harmonic function} on
$D_0$ (if a function satisfies just (ii) and (iii) and is $C^2$ on
$D_0$ we will call it {\it regular}). For such a function we may take
$K = D_0$ in (\ref{Green}) because the integral at the boundary ``at
infinity'' can be easily shown to vanish. Thus, denoting the boundary
data of $u$ on $\Sigma_0$ as $u |_{\Sigma_0} \equiv f_0$ and
$\vec{n}_{\gamma} (u) |_{\Sigma_0} \equiv f_1$ of $u$, the expression
above becomes
\begin{equation}
\int_{\Sigma_0}\left [ \frac{}{} \psi f_1  - f_0  \vec{n}_{\gamma} ( \psi ) \right] \mathrm{d} S_{\gamma} =  0.
\label{GreenBoun}
\end{equation}
These are obviously necessary conditions that the overdetermined
boundary data must satisfy in order for a regular solution $u$ of the
Laplace equation to exist.  It is natural to ask whether such
conditions are also sufficient. More precisely, give two arbitrary
continuous functions $f_0$ and $f_1$ on $\Sigma_0$ which satisfy
(\ref{GreenBoun}) for {\it any} choice of regular $\gamma$-harmonic
function $\psi$. We want to check whether there always exists a
function $u$ satisfying the Laplace 
equation $\triangle_{\gamma} u =
0$, with $r u$ bounded at infinity and such that the boundary
equations $u |_{\Sigma_0} = f_0$, $\vec{n}_{\gamma} (u) |_{\Sigma_0} = f_1$
are satisfied. The answer is yes as we show next. Consider the
Dirichlet problem $\triangle_{\gamma} u = 0 $ with $u |_{\Sigma_0} =
f_0$. Standard elliptic theory tells us that this problem always
admits a unique solution $u$ which tends to zero at infinity. Let us
define $\tilde{f}_1$ on $\Sigma_0$ as $\tilde{f}_1 \equiv
\vec{n}_{\gamma} (u) |_{\Sigma_0}$.  Since $u$ solves the Laplace 
equation, expression (\ref{GreenBoun}) must hold replacing
$f_1$ with 
$\tilde{f}_1$. Furthermore, our assumption is that $f_0$ and $f_1$
satisfy (\ref{GreenBoun}) for any regular $\gamma$-harmonic function
$\psi$. Subtracting both expressions we get $\int_{\Sigma_0} \psi ( f_1 -
\tilde{f}_1 ) \mathrm{d} S_{\gamma} =0$. However, since the regular
$\gamma$-harmonic function $\psi$ is arbitrary and the Dirichlet
problem for the Laplace equation always admits a solution, we have
that $\psi |_{\Sigma_0}$ is an arbitrary continuous function. This
readily implies $f_1 = \tilde{f}_1$ and hence compatibility of the
overdetermined boundary data.

However, condition (\ref{GreenBoun}) has a serious practical
disadvantage, namely that it must be checked for an {\it arbitrary}
decaying solution $\psi$ of the Laplace equation. This makes it not
useful in practical terms. Our aim is to reduce the number of
solutions $\psi$ that must be checked in (\ref{GreenBoun}) while still
implying compatibility of $f_0$ and $f_1$.  Here is where axial
symmetry plays an essential role. This allows us to reduce the number
of conditions to be checked to just a one-parameter family
set. Without going into the details, let us just state that, in the
axially symmetric case, conditions (\ref{GreenBoun}) restricted to the
set of functions
\begin{equation}
\psi_y(\rho,z) = 1/\sqrt{\rho^2 + (z -y )^2},  
\label{psiy}
\end{equation}
where $y$ is a constant such that the point $\{\rho=0, z=y, \phi
\}$ lies {\it outside} the domain $D_0$, are already necessary and
sufficient for existence of the solution $u$. Thus the number of
conditions reduces dramatically and becomes manageable. The main
result we present in this contribution is the generalization of this
result to the four cases corresponding to the perturbed Ernst
equations. The integrals to be performed are, of course, more
complicated in general but the idea is still the same. In order to
write down our main theorem we need to introduce some notation. First
of all, define $z_S < z_N$ as the values of $z$ at the intersection of
$\Sigma_0$ with the axis of symmetry (i.e. the south and north poles
respectively) and restrict $y$ to the interval $(z_S,z_N)$. We
define an angle $\Upsilon_y\in [0, 2 \pi )$ by $\cos \Upsilon_y(\rho,z) \equiv
(z-y)/\sqrt{\rho^2+(z-y)^2}$, $\sin \Upsilon_y(\rho,z) \equiv
\rho/\sqrt{\rho^2+(z-y)^2}$ and three functions
$W_{y}$, $Q_{+}$ and $Q_{-}$ as the unique vanishing at infinity 
solutions of the following compatible PDE 
\begin{eqnarray*}
\mathrm{d} W_{y} & = & 
\cos \Upsilon_y\,\mathrm{d} U_0 + \sin \Upsilon_y\,\star \mathrm{d} U_0, \nonumber \\
\mathrm{d} Q_{\pm }  & = &
  e^{-2 U_0 \pm 2 W_{y} } \left [ \left (  \mp 1 - \cos \Upsilon_y\right )
    \mathrm{d} \Omega'_0 - \sin \Upsilon_y\star \mathrm{d} \Omega'_0 
\right ]. 
\end{eqnarray*}
Here $\star$ means Hodge dual with respect to the metric $dz^2+d\rho^2$.
Then, we have
\begin{theorem}
\label{casebycase}
Let $f_0$, $f_1$ be continuous axially symmetric functions on a $C^1$
simply connected, axially symmetric surface $\Sigma_0$ of $\mathbb{E}^3$.
Let this surface be defined in cylindrical coordinates by $\{ \rho =
\rho_0 (\mu), z = z_0 (\mu), \phi = \varphi \}$, where
$\mu\in [\mu_S,\mu_N ]$ 
and $\mu_S < \mu_N$ are the only solutions
of $\rho_0(\mu)=0$. Call $z_S \equiv z (\mu_S)$ and $z_N \equiv
z(\mu_N)$ and assume $z_S < z_N$ (i.e. that these values correspond to
the ``south'' and ``north'' poles of the surface, respectively).
Denote by $D_0$ the exterior region of this surface and $\vec{n} = -
\dot{z}_0 \partial_{\rho} + \dot{\rho}_0 \partial_z$ a normal vector
to it. Then,

\noindent (i) the Cauchy boundary value problem $
\triangle_{\gamma} U'_0 =0, U'_0 |_{\Sigma_0} = f_0, \vec{n} \left (U'_0 \right ) |_{\Sigma_0} =f_1$
admits a regular solution on $D_0$ if and only if
$$
 \int_{\mu_S}^{\mu_N}\left[\psi_{y} \, f_1 - f_0 \,\vec n(\psi_{y})
 \right]\rho_0 \mathrm{d} \mu = 0, \hspace{2cm} \forall  y \in (z_S,z_N),
$$
(ii) the Cauchy boundary value problem
$\triangle_{\gamma}\Omega'_0- 4\left(\mathrm{d} \Omega'_0,\mathrm{d} U_0\right)_\gamma=0, \Omega'_0|_{\Sigma_0} = f_0,
\vec{n} \left ( \Omega'_0 \right ) |_{\Sigma_0} = f_1$
admits a regular solution on $D_0$ if and only if
$$
 \int_{\mu_S}^{\mu_N}\left[\Psi_{y} \, f_1 - f_0 \,\vec n(\Psi_{y})
 \right]\rho_0 e^{-4 U_0 |_{\Sigma_0}} \mathrm{d} \mu  = 0, \hspace{2cm} \forall  y \in (z_S,z_N),
$$
(iii) the Cauchy boundary value problem
$\triangle_{\gamma} U''_0+e^{-4U_0}
      \left(\mathrm{d} \Omega'_0,\mathrm{d} \Omega'_0\right)_\gamma=0, U''_0|_{\Sigma_0} = f_0,
 $ $\vec{n} \left ( U''_0 \right ) |_{\Sigma_0} = f_1$
admits a regular solution on $D_0$ if and only if
$$
 \int_{\mu_S}^{\mu_N}\left[\psi_{y} \, f_1 - f_0 \,\vec n(\Psi_{y})
- \mathsf{T}_1 \left ( \vec{n} \right )
 \right]\rho_0 \mathrm{d} \mu  = 0, \hspace{2cm} \forall  y \in (z_S,z_N),
$$
and (iv) the Cauchy boundary value problem
$\triangle_{\gamma}\Omega''_0-8
      \left(\mathrm{d} \Omega'_0,\mathrm{d} U'_0\right)_\gamma
      -4\left(\mathrm{d} \Omega''_0,\mathrm{d} U_0\right)_\gamma=0,
\Omega''_0|_{\Sigma_0} = f_0,
\vec{n} \left ( \Omega''_0 \right ) |_{\Sigma_0} = f_1$
admits a regular solution on $D_0$ if and only if
$$
 \int_{\mu_S}^{\mu_N}\left[ \left ( \Psi_{y} \, f_1 - f_0 \,\vec n(\Psi_{y}) \right ) e^{-4 U_0 |_{\Sigma_0}}
-  \mathsf{T}_2  \left (\vec{n}\right )
 \right]\rho_0 \mathrm{d} \mu  = 0, \hspace{2cm} \forall  y \in (z_S,z_N),
$$
where $\psi_y$ is given in (\ref{psiy}), $\Psi_y \equiv 
\frac{e^{ 2 U_0 - 2 W_{y}}}{\sqrt{ \rho^2 + (z - y)^2}}$, 
$\mathsf{T}_1 \equiv \frac{1}{2 \rho} Q_{+} \star \mathrm{d} Q_{-}$ and
$\mathsf{T}_2 \equiv  \frac{8}{\rho} Q_{-} \star
\mathrm{d} \left ( W'_{y}+ U'_0 \right)$.
\end{theorem}

\section*{Acknowledgements}                                                                                
The authors thank EPSRC for funding project GR/R53685/01.
RV thanks the IRCSET 
postdoctoral fellowship Ref. PD/2002/108.

%
%
%

\end{document}